\documentclass[12pt]{utarticle}
\usepackage{amssymb, amsmath, amsopn, amsthm,amscd,amsfonts, fontenc, times, mathptmx, graphicx}
\usepackage{color}
\usepackage{latexsym}
\usepackage{ifthen}
\usepackage{epsfig}
\usepackage{graphicx}
\usepackage{graphics}
\usepackage{bm,longtable,enumerate}

\newcommand{\Tr}{{\rm Tr}}
\newcommand{\tr}{{\rm tr}}

\newcommand{\ol}{\overline}
\newcommand{\del}{\partial}

\newcommand{\nn}{\nonumber}

\newcommand{\eqn}{\begin{eqnarray}}
\newcommand{\eqnx}{\end{eqnarray}}

\newcommand{\bZ}{\mathbb{Z}}
\def\beq{\begin{equation}}
\def\eeq{\end{equation}}
\def\beqa{\begin{eqnarray}}
\def\eeqa{\end{eqnarray}}
\def\scs{\scriptscriptstyle}

\def\tr{\mathop{\rm tr}\nolimits}
\def\Tr{\mathop{\rm Tr}\nolimits}

\newcommand{\bH}{\mathbb{H}}
\newcommand{\ba}{{\boldsymbol{a}}}

\newcommand{\bk}{{\vec{k}}}
\newcommand{\bp}{{\vec{p}\,}}
\newcommand{\bX}{{\vec{X}}}

\newcommand{\psl}{\not\! p}

\def\matt[#1,#2,#3,#4]{\left(%
\begin{array}{cc} #1 & #2 \\ #3 & #4 \end{array} \right)}

\def\v2#1{\vv2[#1]}
\def\vv2[#1,#2]{\left(%
{#1 \atop #2}\right)}

\def\ol{\overline}

\def\nn{\nonumber}

\def\eps{\epsilon}

\title{\bf Towards Relativistic Skyrmions}

\begin{document}
\author{ {\sc Henrique Boschi-Filho $^{\rm a}$ , Nelson R.~F.~Braga} 
 \address{
 Instituto de F{\'i}sica, \\
 Universidade Federal do Rio de Janeiro,\\
         Caixa Postal 68528,\\
         21941-972 Rio de Janeiro, RJ, Brasil.\\[0.2cm]
         Email: {\tt boschi@if.ufrj.br},\\
         ~~~~~~~~~{\tt braga@if.ufrj.br},\\
         ~~~~~~~~~{\tt mtorres@if.ufrj.br}
          } ,
 {\sc Matthias Ihl}
 \address{
        School of Theoretical Physics, \\
        Dublin Institute for Advanced Studies, \\
        10 Burlington Rd, \\
        Dublin 4, Ireland. \\[0.3cm]
        Email: {\tt msihl@stp.dias.ie}
         } ,
 \\ {\sc Marcus A.~C.~Torres} $^{\rm a}$  
}

\Abstract{We revisit baryons in the Skyrme model. Starting from static baryons in the helicity eigenstates, we generalize their wavefunctions to the non-static and relativistic regime. A new representation for gamma matrices in the soliton collective space is constructed and the corresponding Dirac equation is obtained. As an example, we draw consideration on how to apply this new representation on the calculus of vector current vacuum expectation values for baryon states of spin and isospin half and arbitrary momenta and we show how elastic form factors can be derived.\\
7th Conference Mathematical Methods in Physics - Londrina 2012, Rio de Janeiro, Brazil}
		
\maketitle
\newpage

\section{Introduction}

 Skyrmions are solitons, classical, static solution from NL$\sigma$M of pions \cite{Skyrme:1961vq} and they are identified as baryons in such model. It is a model with chiral Lagrangean and spontaneous chiral symm breaking. It has attracted renewed interest as a simplified description of baryons in the large $N_c$ limit of QCD. Here we review the work presented in \cite{BoschiFilho:2011hn}   where we show a simple way to overcome its static or slow moving condition, extending it to a relativistic framework.

The static properties of baryons in the Skyrme model (skyrmions) have been studied in \cite{Adkins:1983ya} , where baryonic quantum states emerge from a canonical quantization of the soliton moduli.
In \cite{Sakai:2004cn}, 
Sakai and Sugimoto derive the Skyrme model as the $3+1$ dimensional pion effective theory descending from the dynamics of the gauge fields living on flavor $D8$-branes that probe
the $D4$-brane geometry generated by a stack of color branes. 

Here, after introducing the Skyrme model and the soliton canonical quantization in section 2, we calculate relativistic baryon wave functions in their helicity eigenstates (section 3) generalizing the results of \cite{Adkins:1983ya}  by boosting  wave functions in the momentum direction. We show that the state representation is in a form compatible with the Dirac spinor representation. In section 4, as an example we show how to obtain  form factors from the Lorentz tensorial decomposition of  matrix elements of baryonic currents in our relativistic approach.

\section{Skyrme Model}

Skyrmions are soliton solutions of a nonlinear effective field theory of pions.
The action of the Skyrme model is given by
\begin{eqnarray}
S=\int d^4 x\left(\frac{f_\pi^2}{4}\tr\left(
 U^{-1}\del_\mu U\right)^2+\frac{1}{32 e^2}
\tr\left[U^{-1}\del_\mu U,U^{-1}\del_\nu U\right]^2
\right) \, 
\end{eqnarray}
where the pion fields $\pi(x^\mu)$ are encoded in the SU(2) valued Skyrme field,
\beq
\;\;U(x^\mu)= \frac{\sigma(x^\mu) \mathbb I +\pi^i(x^\mu)\cdot \tau^i}{f_\pi} =e^{i\pi(x^\mu)/f_\pi} {\rm \tiny  (weak\, pion\, field\, limit)},  \nn
\eeq 
where $\sigma^2(x^\mu)+ \vec\pi^2(x^\mu)=f_{\pi}^2$. U transforms as
$\;\;U(x^\mu)\rightarrow g_L U(x^\mu)  g_R^{-1}\;\;$ under  the chiral symmetry $ SU(2)_L\times SU(2)_R$ present in the Lagrangian.

The Skyrmion is a static solution that goes to unity value at infinity breaking chiral symmetry to the diagonal subgroup. In this case U becomes a map
\beq U:{\mathbb R}^3+\{\infty\}= S^3\rightarrow SU(2)\equiv S^3
\eeq
and a topological charge $n_B \in \pi_3(SU(2))= {\mathbb Z}$ is identified with 
the baryon number.

\subsection{SU(2) collective coordinates}

We begin by reviewing the moduli space approximation method to quantize Skyrmions. The fundamental idea \cite{Adkins:1983ya} is to realize that there is a moduli of soliton solution given by  $SU(2)$ rotations of a given one, $U_0$, and define a slowly moving soliton by a  time dependent rotation, 
\beq
U(t,x^M)=VU_0(x^M,X^M(t))V^{-1},
\label{su2moduli}
\eeq
where $V=V(t,x^M)$ is an $SU(2)$ element and $X^M$, $M=\{1,\ldots, 3\}$, represents the position of the soliton in the spatial $\mathbb{R}^3$. Introducing the collective coordinates $\ba(t)=a_4(t)+ia_a(t)\tau^a$ as a point in $S^3$ representing the $SU(2)$ orientation,  we note $V(t,x)\rightarrow \ba(t)\; {\rm as}\;x\rightarrow \infty$. 

 When   $\ba(t) U_o\ba^{-1}(t)$ is replaced in the original Lagrangian, a new Lagrangian emerge, 
\beq
L=-M+2\lambda\sum_{i=1}^{4}(\dot{a_i})^2
\eeq

Introducing conjugate momenta $\; \pi_i=\partial L/\partial \dot{a_i}= -i\frac{\partial}{\partial a_i}$, we find the Hamiltonian: 
\beq\;\; H = M+ \lambda\nabla^2_{S^3}\eeq

The eigenstates (wavefunctions) are factorized into radial and spherical harmonics components. On the $S^3$, they are scalar spherical harmonics and are known to be homogenous polynomials
\beq
T^{(l)}(a_I)=C_{I_1\cdots I_l}\,a_{I_1}\cdots a_{I_l}\ ,
\eeq
where $C_{I_1\cdots I_l}$ is a traceless symmetric tensor of rank
$l$. 

The dimension of the tensor $C$ and therefore the dimension of the space of spherical harmonics of degree $l$, $\bH_l$ is $(l+1)^2$. The space $\bH_l$ is a representation of the rotation group $SO(4)$ and corresponds  to the $(S_{l/2},S_{l/2})$ representation of
$(SU(2)\times SU(2))/\bZ_2\simeq SO(4)$.  $S_{l/2}$ denotes the spin $l/2$
representation of $SU(2)$, with $\dim S_{l/2}= l+1$.
Under  group rotation $\ba$ transforms as
\begin{equation}
\ba \to g_I\,\ba\, g_J \ ,\quad
g_{I,J}\in SU(2)_{I,J} \ .
\label{gag}
\end{equation}
where $SU(2)_I$ and $SU(2)_J$ are identified with the
isospin rotation and the spatial rotation, respectively. 
\section{Static and Relativistic Baryons}
The  collective state is quantized considering slowly moving solitons. We will extend it to relativistic baryons by simply treating them as static solitons boosted in a given direction. First, we review the spherical harmonics tensors to static nucleons. The lowest states are at $l=1$ and the tensors become linear in $a_I$ coordinates. They correspond to states with spin and isospin $1/2$ and we identify them with protons and neutrons. In spinorial notation we write the particle states as
\beq
|N,h\rangle= \chi^N \otimes\chi_h =: \chi^N_h,
\eeq
where $N=\{p,n\}$, $h=\{+,-\}$ and
\beq
\chi^{p}=\chi_{+} =\left(\begin{array}{c}1\\0\end{array}\right) \quad , \quad \chi^{n}=\chi_{-} =\left(\begin{array}{c}0\\1\end{array}\right).
\label{spinorial notation}
\eeq
The isospin $I_3$ and spin $J_3$ operators in this representation read
\begin{eqnarray}
I_a=\frac{i}{2}\left(
a_4\frac{\del}{\del a_a}-a_a\frac{\del}{\del a_4}
-\epsilon_{abc}\,a_b\frac{\del}{\del a_c}
\right)\ ,~~
J_a=\frac{i}{2}\left(
-a_4\frac{\del}{\del a_a}+a_a\frac{\del}{\del a_4}
-\epsilon_{abc}\,a_b\frac{\del}{\del a_c}
\right),
\label{IJ}
\end{eqnarray}
and their eigenstates are given by
\beq
|p,+\rangle=\frac{a_1 + i a_2}{\pi}\;, \; |p,-\rangle=-\frac{i(a_4-i a_3)}{\pi}\;,\; 
|n,+\rangle=\frac{i(a_4 + i a_3)}{\pi} \;, \; |n,-\rangle=-\frac{(a_1-i a_2)}{\pi}. 
\label{nucleonwavefunctions}
\eeq

\subsection{Nucleon relativistic wavefunctions}
 In generalizing the expressions aboveto the relativistic case, we initially restrict the discussion to proton and neutron states separately and disregard the isospin information. 
We introduce a relativistic spinor for a fermion with a given momentum $\vec p$, 
\beqa
u(p,h) = \frac{1}{\sqrt{2E}} \left( \begin{array}{c} f \chi_h \\ \frac{\vec{p} \cdot \vec{\sigma}}{f} \chi_h \end{array} \right)\quad,\; {\rm with}  \; f = \sqrt{E + m_B} .
\label{u}
\eeqa
Here $\chi_h$ is a helicity eigenstate, which means $\vec{p} \cdot \vec{\sigma}\chi_h=h |\vec p| \chi_h$ where $h= \pm 1$. 
This yields the eigenstates
\beq
\chi_{+}(\vec p) =\frac{1}{\sqrt{2|\vec p|(|\vec p|+p_3)}}\left(\begin{array}{c}|\vec p|+p_3\\p_1+i p_2\end{array}\right) \quad , \quad \chi_{-}(\vec p) =\frac{1}{\sqrt{2|\vec p|(|\vec p|+p_3)}}\left(\begin{array}{c}-p_1+i p_2\\|\vec p|+p_3\end{array}\right).  
\label{helicity spinors}
\eeq
Notice that  $\chi_h^{\dagger}\chi_h=1$ and $|\vec p|= \sqrt{p_1^2+p_2^2+p_3^3}$.
Hence we expect the proton and neutron linear ($l=1$) spherical harmonic tensors (equivalent to $\chi_h(\vec{p})$ helicity eigenstates) to be
\begin{align}
\chi_+^{p}(a_i,\vec{p})\hspace{-.1cm}&=&\hspace{-.4cm} \frac{(|\vec p|\hspace{-.1cm}+\hspace{-.1cm}p_3)(a_1\hspace{-.1cm}+\hspace{-.1cm}i a_2)\hspace{-.1cm}-\hspace{-.1cm}i(p_1\hspace{-.1cm}+\hspace{-.1cm}i p_2)(a_4\hspace{-.1cm}-\hspace{-.1cm}i a_3)}{\pi\sqrt{2|\vec p|(|\vec p|+p_3)}},\; 
\chi_-^{p}(a_i,\vec{p})\hspace{-.1cm}=\hspace{-.1cm} \frac{(\hspace{-.1cm}-p_1\hspace{-.1cm}+\hspace{-.1cm}i p_2)(a_1\hspace{-.1cm}+\hspace{-.1cm}i a_2)\hspace{-.1cm}-\hspace{-.1cm}i(p_3\hspace{-.1cm}+\hspace{-.1cm}|\vec p|)(a_4\hspace{-.1cm}-\hspace{-.1cm}i a_3)}{\pi\sqrt{2|\vec p|(|\vec p|+p_3)}},\nn\\
\chi_+^{n}(a_i,\vec{p})\hspace{-.1cm}&=&\hspace{-.9cm} \frac{(|\vec p|\hspace{-.1cm}+\hspace{-.1cm}p_3)i(a_4\hspace{-.1cm} +\hspace{-.1cm} i a_3)\hspace{-.1cm}-\hspace{-.1cm}(p_1\hspace{-.1cm}+\hspace{-.1cm}i p_2)(a_1\hspace{-.1cm}-\hspace{-.1cm}i a_2)}{\pi\sqrt{2|\vec p|(|\vec p|+p_3)}},\;
\chi_-^{n}(a_i,\vec{p})\hspace{-.1cm}=\hspace{-.1cm} \frac{(\hspace{-.1cm}-p_1\hspace{-.1cm}+\hspace{-.1cm}i p_2)i(a_4\hspace{-.1cm} +\hspace{-.1cm} i a_3)\hspace{-.1cm}-\hspace{-.1cm}(p_3\hspace{-.1cm}+\hspace{-.1cm}|\vec p|)(a_1\hspace{-.1cm}-\hspace{-.1cm}i a_2)}{\pi\sqrt{2|\vec p|(|\vec p|+p_3)}}.
\label{helicity wavefunctions}
\end{align}
The expression above is not valid for $p_1=p_2=0$ and $p_3 = -p$ and in that case we recall that helicity changes sign when momentum changes in sign to
 the opposite direction,
\beq
\chi_h^{N}(a_i,-\vec{p})\equiv \chi_{-h}^{N}(a_i,\vec{p}),
\label{opposite direction}
\eeq

After some algebra we can verify that (\ref{helicity wavefunctions}) are eigenfunctions of $\bp. \vec J$ operators (\ref{IJ}):
\beq
\bp.\vec J\,\chi_+^{N}(a_i,\vec{p})=+\frac{1}{2}|\vec p| \,\chi_+^{N}(a_i,\vec{p})\quad {\rm and} \quad \vec p.\vec J\,\chi_-^{N}(a_i,\vec{p})=-\frac{1}{2}|\vec p| \,\chi_-^{N}(a_i,\vec{p}).
\eeq
Similarly to Dirac 4-spinors (\ref{u}),The relativistic nucleon SU(2) wavefunctions are
\beq
u_N(a_i,\bp,+)= \frac{1}{\sqrt{2E}} \left( \begin{array}{c} f \chi_+^{N}(a_i,\vec p) \\ \frac{|\vec{p}|}{f} \chi_+^{N}(a_i,\vec p) \end{array} \right) \quad , \quad 
u_N(a_i,\bp,-) = \frac{1}{\sqrt{2E}} \left( \begin{array}{c} f \chi_-^{N}(a_i,\vec p) \\ -\frac{|\vec{p}|}{f} \chi_-^{N}(a_i,\vec p)  \end{array} \right) \, .
\eeq
The explicit dependence on $a_I \in S^3$ is not directly observable and the integration in $S^{3}$ moduli recovers the 4D Dirac spinor with appropriate isospin.

\subsection{Dirac equation and spin sum}
 
In order to work with relativistic $SU(2)$ wavefunctions instead of Dirac spinorial notation, we define the substitutes of gamma matrices in the $SU(2)$ collective space by simply replacing the Pauli matrices $\sigma^i$ with $2 J^i$ operators (\ref{IJ}).  Hence, the new $2\times2$ gamma matrices are
\beq
\gamma^0=-i\left( \begin{array}{cc} 1&0 \\ 0&-1 \end{array} \right) \quad , \quad 
\gamma^i=-i\left( \begin{array}{cc} 0&2J^i \\ -2J^i &0 \end{array} \right)  \, .
\eeq
Such operators act only on spin and we will disregard the isospin index for now. Upon such substitution we can verify the validity of the Dirac equation:
\beqa
(i  \!\psl+m_B)u_N(a_i,\bp,h)=0.\\
\ol u_N(a_i,\bp, h) (i  \!\psl+m_B)= 0.
\eeqa
Since $J^i$ operators have real eigenvalues and behave like Pauli matrices, we define their operation to the left by transpose conjugation:
$\psi_h^\dagger (\vec p. \vec J)=(\vec p . \vec J \psi_h)^\dagger = |\vec p| \frac{h}{2} \psi^\dagger_h$.

As mentioned before, the spinor normalization is given by an integration of the $a_i$ moduli,
\beq
\ol u(\bp, h')u(\bp, h)= \int_{S^3}\ol u(a_i,\bp, h')u(a_i,\bp, h)
\eeq
Working out the integrand,
\beqa
 \ol u(a_i,\bp, h')u(a_i,\bp, h)&=& \frac{1}{2E}(f\chi_{h'}^{*}(a_i,\bp)\quad \tfrac{|\bp|}{f}h'\chi_{h'}^*(a_i,\bp))\left( \begin{array}{cc} 1&0 \\ 0&-1 \end{array} \right) \left( \begin{array}{c} f \chi_h^{a_i}(\vec p) \\ \frac{|\vec{p}|}{f}h \chi_h(a_i,\vec p) \end{array} \right) \nn\\
&=& \frac{1}{2E}\left(f^2-h'h\frac{|\bp|^2}{f^2}\right)\chi_{h'}^{*}(a_i,\bp)\chi_h(a_i,\bp)
\label{oluuintegrand}
\eeqa
In order to integrate (\ref{oluuintegrand}) over $S^3$ we write $(a_1,a_2,a_3,a_4)$ in spherical coordinates:
\beq
a_1=\sin\theta_0\sin\theta_1\sin\theta_2, \; a_2=\sin\theta_0\sin\theta_1\cos\theta_2,\;
a_3=\sin\theta_0\cos\theta_1,\;a_4=\cos\theta_0.\nn
\eeq
The volume element is $d\Omega_3 = \sin^2\theta_0 \sin\theta_1
d\theta_0 d\theta_1 d\theta_2$. Using (\ref{helicity wavefunctions}), the integration over (\ref{oluuintegrand}) turns out to be
\beq
\int_{S^3} \ol u(a_i,\bp, h')u(a_i,\bp, h) =\frac{1}{2E} \left((E+m_B)-h'h(E-m_B)\right)\int_{S^3}\chi_{h'}^{a_i*}(\bp)\chi_h^{a_i}(\bp)=\frac{m_B}{E}\delta_{h'h}.
\label{unorm}
\eeq
The spin sum is given by
\beq
\sum_hu(\bp,h)\ol u(\bp,h)=\sum_h\int_{S^3}u(a_i,\bp,h)\ol u(a_i,\bp,h), 
\label{spinsum}
\eeq
where the integrand of (\ref{spinsum}) reads
\beqa
u(a_i,\bp,h)\ol u(a_i,\bp,h) &=&  \frac{1}{2E}\left( \begin{array}{c} f \chi_h^{a_i}(\vec p) \\ \frac{2\vec J.\bp}{f}\chi_h^{a_i}(\vec p) \end{array} \right)(f\chi_h^{a_i*}(\bp)\quad -\tfrac{2\vec J.\bp}{f}\chi_h^{a_i*}(\bp)) \nn\\
= \frac{1}{2E}&&\hspace{-.5cm}\left(\begin{array}{cc}E+m_B&-2\vec J.\bp\\2\vec J.\bp&-E+m_B\end{array}\right)|\chi_h^{a_i}(\bp)|^2  = \frac{1}{2E}(-i  \!\psl+m_B)|\chi_h^{a_i}(\bp)|^2.
\label{spinsumintegrand}
\eeqa
Integrating over (\ref{spinsumintegrand}) we get the spin sum (\ref{spinsum}),
\beqa
\sum_h\int_{S^3}u(a_i,\bp,h)\ol u(a_i,\bp,h)=\frac{1}{2E}(-i  \!\psl+m_B)\sum_h\int_{S^3}|\chi_h^{a_i}(\bp)|^2=\frac{1}{E}(-i  \!\psl+m_B).
\eeqa

\section{Application:  elastic form factors}

  The SU(2) vector current integrated over the unit sphere of spatial $\mathbb R^3$ \cite{Adkins:1983ya} behave as
\beq
\int d \Omega^2 J^{0,c}_V\sim \; \Tr[(\partial_0\ba)\ba^{-1}\tau^c]\, , \quad {\rm and} 
\quad \eps^{ijk}\hspace{-2mm}\int d \Omega^2 x^j J^{k,c}_V\sim \; \Tr[\tau^i \ba^{-1}\tau^c\ba],
\eeq
where it should be noted that the $SU(2)_V$ currents were decomposed as $J^\mu_V=J^{\mu,c}_V\tau^c$.
Since the collective coordinates do not depend on the position in $\mathbb R^3$, general vector currents read
\beq
  J_V^{0,c}(\bk)=e^{-i\bk\cdot\bX}I_c\;,\;
  J_V^{i,c}(\bk)=ie^{-i\bk\cdot\bX}\,\Lambda
\epsilon_{ija}q_j\Tr\left(\tau^c\ba\tau^a\ba^{-1}\right)
\label{wtJj}
\eeq
where $q_j := p'_j - p_j$. 
When calculating the vector current proton-proton matrix elements, we are only interested in the $c=3$ components (see above). 
We define the Dirac and Pauli form factors according to the following decomposition of current matrix elements, 
\beq \label{currentdecomp}
\hspace{-4mm}\langle p_{\scs  X} , B_{\scs  X} , s_{\scs  X}  \vert J_V^{\mu,a} (0) \vert p , B, s \rangle \hspace{-1mm}=\hspace{-1mm} \frac{i(\tau^a)_{\scs I_3^{\scs  X} I_3} }{2 (2 \pi)^3}  
( \eta^{\mu \nu}  - \frac{ q^\mu q^\nu}{q^2} )
 \bar u (p_{\scs  X}  , s_{\scs  X})\hspace{-1mm} \Big [  \gamma_\nu F^{D,a}_{B B_{\scs  X}}(q^2) 
 + \kappa_B  \sigma_{\nu \lambda} q^\lambda  F^{P,a}_{B B_{\scs  X}} (q^2)\hspace{-1mm}  \Big ]\hspace{-1mm} u (p , s)
\eeq
In the elastic case, in the Breit frame, with $\bp=-\bp'=-\dfrac{q}{2} \hat z$, the expression (\ref{currentdecomp}) becomes 
\beq
 \hspace{-3mm}\langle p_{\scs  X} , B_{\scs  X} , s_{\scs  X}  \vert J_V^{0,a} (0) \vert p , B, s \rangle
\hspace{-1mm}=\hspace{-1mm} \frac{1}{2 (2 \pi)^3}  (\tau^a)_{I_3^{\scs  X} I_3} \left ( \frac{m_B}{E} \right ) \delta_{s_{\scs X},- s}
\left [ F^{D,a}_{B}(q^2) - \frac{q^2}{4 m_B^2} F^{P,a}_{B}(q^2) \right ]
\label{sachse}
\eeq
\beq
\hspace{-7mm}\langle p_{\scs  X} , B_{\scs  X} , s_{\scs  X}  \vert J_V^{i,a} (0) \vert p , B, s \rangle
=  -\frac{1}{2 (2 \pi)^3}  (\tau^a)_{I_3^{\scs  X} I_3} \left ( \frac{i}{2E} \right )
\epsilon^{ijk} q_j (\sigma_k)_{s_{\scs  X},- s}
\left [ F^{D,a}_{B}(q^2) + F^{P,a}_{B}(q^2) \right ]
\label{sachsm}
\eeq
The equations above contain a relativistic correction not present in Skyrme Models. \\
We need to calculate the  vevs of  $\Tr(\tau^a\ba^{-1}\tau^3\ba)$ in (\ref{wtJj}) in order to find the form factors.

In spherical coordinates the traces become (a=3):
\beqa
\Tr(\tau^3\ba^{-1}\tau^3\ba)&=&2 (\cos(\theta_0)^2 + \cos(2 \theta_1) \sin(\theta_0)^2).
\eeqa
The general result for  baryon states  (a=1,2,3) is
\begin{align}
\langle  \Tr(\tau^a\ba^{-1}\tau^3\ba)\rangle\equiv\langle B_X, \bp', h' |\Tr(\tau^a\ba^{-1}\tau^3\ba)|B_0,\bp,h \rangle=\hspace{1cm}\nn\\
\int_{S^3}\hspace{-.3cm}d\Omega^3\;\ol u_X(a_i,\bp',h')\Tr(\tau^a\ba^{-1}\tau^3\ba)u_0(a_i,\bp,h)
\hspace{-.1cm}=\hspace{-.1cm}\tfrac{-1}{3\sqrt{E_XE}}(ff_X\hspace{-.1cm}-\hspace{-.1cm}\tfrac{hh'|\bp||\bp'|}{ff_X})\chi^\dagger_{h'}(\bp')\sigma^a\chi_h(\bp)\hspace{.4cm}
\label{general trace}
\end{align}

where $f_X=\sqrt{E_X+m_{B_X}}$ and $\chi_{h'}(\bp')$ and $\chi_h(\bp)$ are the helicity spinors (\ref{helicity spinors}). 
The equation above plays an important role in the calculation of the electromagnetic proton form factors \cite{Sakai:2004cn}. In the elastic case, in Breit frame, $\bp=-\bp'=-\dfrac{q}{2}\hat z$ and (\ref{general trace}) takes the form
\beq
\langle \Tr(\tau^a\ba^{-1}\tau^3\ba)\rangle = \frac{(E+m_B-h'h(E-m_B))}{2E}(-\frac{2}{3})\sigma^a_{h', -h} \hspace{-1mm} =\hspace{-2mm}\left\{ \;\begin{array}{l} \; -\frac{2}{3} \frac{m_B}{E}\sigma^1_{h', -h}\; {\rm for }\; a=1.\\ \;-\frac{2}{3} \frac{m_B}{E}\sigma^2_{h', -h}\; {\rm for }\; a=2.\\ \;-\frac{2}{3}\sigma^3_{h', -h}\; \;{\rm for }\; a=3.  \end{array}\right.
\label{zcorrection}
\eeq
where we used the substitution (\ref{opposite direction}). This amounts to a relativistic correction in the direction of movement $\hat z$ w.r.t.~the static case \cite{Adkins:1983ya},  of a factor of $\frac{m_B}{E}$.

Using  (\ref{zcorrection}) in the vector currents (\ref{wtJj}) , the current matrix elements can be written as
 \beqa
\langle \dfrac{q}{2} \hat z, B_{\scs  X} , h'  \vert J_V^{0,3} (0) \vert-\dfrac{q}{2} \hat z, B, h \rangle &=& 
\frac{1}{2 (2 \pi)^3} \left(\frac{m_B}{E}\right) \delta_{-h'h}  \label{Je0}\\
\langle \dfrac{q}{2} \hat z , B_{\scs  X} , h'  \vert J_V^{i,3} (0) \vert -\dfrac{q}{2} \hat z , B, h \rangle &=& -\frac{1}{2(2 \pi)^3} \left(\frac{i}{2E}\right) \frac{8m_B}{3} \Lambda 
\epsilon_{i3a} q_3 \sigma^a_{-h'h}
\label{Jej}
\eeqa
where we utilized eq.~(\ref{unorm}) and noted that $\langle 2 I^3 \rangle_{+1/2,+1/2}=(\tau^3)_{+1/2,+1/2}=1$ for the proton-proton matrix elements.
Employing eqs.~(\ref{Je0}), (\ref{Jej}) and eqs.~(\ref{sachse}), (\ref{sachsm}) we arrive at
\beq
 F_B^{D,3}(q^2)=\left( 1 + \frac{q^2}{4 m_B^2} \right)^{-1}\hspace{-2mm} \left( \frac{2}{3}\frac{\Lambda q^2}{m_B}  + 1\right),\quad
 F_B^{P,3}(q^2)=\left( 1 + \frac{q^2}{4 m_B^2} \right)^{-1} \hspace{-2mm}\left( \frac{8}{3}m_B \Lambda -1 \right).
\label{formfactorsDP}
\eeq
They are proton form factors for $\tau^3$ isovector current .
\section{Conclusion}
Here we presented a relativistic generalization for non-static baryons in Skyrme models. The correct relativistic normalization for currents and the vacuum expectation values appears 
quite naturally in our approach. 
\section*{Acknowledgements}
The authors H.B-F., N.R.F.B. and M.A.C.T. are supported by CAPES and CNPq (Brazilian research agencies). The work of M.I. was supported by an IRCSET postdoctoral fellowship.

\end{document}